\documentclass[pra,showpacs,amssymb]{revtex4}

\usepackage{amsmath}
\usepackage{amsfonts}
\usepackage{dcolumn}
\usepackage[dvips]{graphicx}

\def\>{\rangle}
\def\<{\langle}

\def\be{\begin{equation}}
\def\ee{\end{equation}}

\begin{document}
\title{Complete eigenstates of N identical qubits arranged in regular polygons}

\author{Terry Rudolph}
\email{rudolpht@bell-labs.com}
\affiliation{Institut f\"ur Experimentalphysik, Universit\"at Wien, Boltzmanngasse 5,
1090 Vienna, Austria}
\affiliation{Bell Labs, 600-700 Mountain Ave., Murray Hill, NJ 07974, U.S.A.}

\author{Itay Yavin}
\author{Helen Freedhoff}
\affiliation{Department of Physics and Astronomy, York University,
Toronto, ON M3J 1P3, Canada}

\date{\today}

\begin{abstract}
We calculate the energy eigenvalues and eigenstates corresponding
to coherent single and multiple excitations of an array of $N$
identical qubits or two-level atoms (TLA's) arranged on the
vertices of a regular polygon. We assume only that the coupling
occurs via an exchange interaction which depends on the separation
between the qubits. We include the interactions between {\it all}
pairs of qubits, and our results are valid for arbitrary distances
relative to the radiation wavelength.  To illustrate the
usefulness of these states , we plot the distance dependence of
the decay rates of the $n=2$ eigenstates of an array of 4 qubits,
and tabulate the biexciton eigenvalues and eigenstates, and
absorption frequencies, line widths, and relative intensities for
polygons consisting of $N=2,\cdots,9$ TLA's, in the
long-wavelength limit.
\end{abstract}


\maketitle

In this paper, we calculate the eigenstates of $N$ identical
(coherently excited) two-level quantum systems arranged on the
vertices of a regular polygon. Such systems are known as {\it
qubits} to the quantum information community and as {\it two-level
atoms (TLAs)} to the quantum optics/spectroscopy community.  The
coherent excitation of identical TLAs has long been of interest to
spectroscopists, in connection with the theory of molecular
excitons \cite{[excitons]} for example, or the phenomenon of
superradiance \cite{[superr]}.  More recently, the interest has
been in connection with the optical properties of molecular
clusters or aggregates \cite{[muk]}, many of which properties are
believed to be related to the coherent interaction of the
aggregates with the radiation field.  At the same time,
multiparticle entangled states of qubits have become an active
area of study in the field of quantum information theory. The
results presented here are of relevance to this community in the
studies of decoherence-free subspaces \cite{[lidar]} and
investigations into the entanglement properties of rings of qubits
\cite{[oconnor]}.\smallskip

We emphasize the complete generality of the majority of the
results obtained herein: our results are applicable to all systems
in which excitation is exchanged between the pairs of interacting
qubits.  Such exchange interactions occur widely: For example, our
theory is applicable to systems in which the coupling is via a
spin-exchange interaction, or via a retarded dipole-dipole
(quadrupole-quadrupole) interaction, such as exists in coherent
dipole \cite{[dicke]} (quadrupole \cite{[hsfI]}) radiative
excitation of atoms or molecules.  We do {\it not} make the common
approximation of including only nearest-neighbour interactions,
but rather we diagonalize the full Hamiltonian, and for arbitrary
distances relative to the radiation wavelength. This is important
for many physically realistic systems, in which the coupling
between non-nearest neighbours can exceed that between adjacent
qubits.\smallskip

The eigenstate calculations are presented in section \ref{calcs}.
In section \ref{calcsA}, we begin by reviewing the calculation of
the eigenstates for single ($n=1$) excitations of a system of $N$
qubits arranged at the vertices of a regular polygon and
interacting via an exchange interaction, valid for arbitrary $N$.
Next, in section \ref{calcsB} we present a method for calculating
the eigenstates for double ($n=2$) excitations of the system, also
for arbitrary $N$. Finally in section \ref{calcsC}, we outline the
calculation of the triplet ($n=3$) eigenstates for $N=6$ and $7$,
and present in Tables I-III the complete set of eigenstates for
all regular polygons up to and including $N=6$; results for $N=7$
are available upon request.\smallskip

 In section \ref{applications}, we present some results specific to the physical
realization of qubits in terms of TLAs interacting via a retarded
dipole-dipole (or quadrupole-quadrupole) interaction.  We have
special interest in the total decay rates of these eigenstates in
order to identify particularly long lived states, which may be
useful in encoding quantum information.  To quantum information
theorists, these are known as ``decoherence-free'' states, and to
spectroscopists as ``subradiant'' states \cite{[subr]}.  In
general, complete subradiance exists only in the small sample
limit, when distance effects are ignored.  Since our calculations
contain the complete distance dependence, they can be used to
examine deviations from the ``long wavelength'' or ``equal
collective decoherence'' assumption commonly made in the theory of
decoherence-free subspaces.\smallskip

 In the spectroscopy community, the study of collective atomic phenomena is many years
old, beginning with Dicke's pioneering article \cite{[dicke]}; for
the early work, see \cite{[superr], [hsfII]}, and references
therein.  A detailed study of the cooperative emission by a
fully-excited system of 3 identical atoms in some specific
geometrical configurations was performed by Richter \cite{[rich]},
while the complete eigenstates for two- and three-atom systems of
arbitrary geometrical arrangement can be found in reference
\cite{[hsfII]}. The single-excitation eigenstates of linear chains
were presented in \cite{[chains]}, of two-dimensional arrays in
\cite{[2d]}, and of rings and regular polygons in \cite{[hsfIII]}.
Single and double excitations of regular polygons in the
long-wavelength limit were considered by Spano and Mukamel
\cite{[muk]}; however, they included in their calculations only
nearest-neighbour interactions, so that our energy eigenvalues and
eigenstates differ considerably from theirs.\smallskip

\section{The Calculations}\label{calcs}

We consider systems of $N$ identical qubits located at positions
${\bf r}_i$, each with ground state $|0_i\rangle$, excited state
$|1_i\rangle$ and transition frequency $\omega$. The free
Hamiltonian is given by $$H_f=\sum_{i=1}^N\hbar \omega
|1_i\rangle\langle 1_i|.$$ Henceforth, we will label states in the
``computational'' basis, i.e. the bare uninteracting states,
according to which atoms are excited therein.  For example, the
state of the $N=5$ system in which atoms 2 and 5 are excited is
written by quantum opticians as $|gegge\rangle$, by quantum
information theorists as $|01001\rangle$, and by us here as
$|25\rangle$. The state with all qubits in the state $|0\rangle$
($|1\rangle$) is denoted by $|G\rangle$ ($|E\rangle$).\smallskip

The generic (excitation-)exchange interaction Hamiltonian of the
qubits is given by \be\label{hint} H_{int}=\sum^N_{{i,j=1} \atop
{i\ne j} }\hbar\Omega_{ij}S_{i}^{+}S_{j}^{-}, \ee where $S_i^+$
and $S_i^-$ are the raising and lowering operators of qubit $i$.
The sole assumption we make regarding the interaction potential
$\Omega_{ij}$ is that  it is a function only of the separation
between qubits $i$ and $j$, ${\bf r}_{ij}={\bf r}_i-{\bf r}_j$. We
focus in this paper on qubits arranged at the vertices of regular
polygons, and number them sequentially around the polygon (see
Fig. 1).  For nearest-neighbour qubits, we define
$\Omega_{i,i\pm1}=a$; similarly, $\Omega_{i,i\pm2}=b$; for $N$
atoms there are $\lfloor N/2 \rfloor$ characteristic interactions,
which we label sequentially alphabetically.\smallskip

In analysing the system of interacting qubits we are faced with
two possibilities; in this paper we follow option (ii):\smallskip

\noindent(i) We can take the $\Omega_{ij}$ to be real, for example
equal to the well known expression for the dipole-dipole
interaction energy. Diagonalizing the interaction Hamiltonian
yields eigenvalues which are the energy level shifts. The dynamics
of the system can then be analyzed using a master equation, which
would include terms containing the dipole-dipole damping. By
solving the master equation we can calculate various quantities of
interest, in particular the total decay rate from a given
eigenstate to all the states below it; this is useful for
identifying long lived states desirable for quantum
computing.\smallskip

\noindent(ii) Alternatively, we can include the free-atom
radiation damping $H_d=\sum_{i=1}^N\hbar\gamma|1_i\rangle\langle
1_i|$, and as well take the $\Omega_{ij}$ to be complex. The real
part of $\Omega_{ij}$ is then the interaction energy, the
imaginary part the inter-qubit damping. Although this results in a
non-Hermitian Hamiltonian, it has the advantage that the imaginary
part of each resulting eigenvalue automatically contains the total
decay rate for that eigenstate (the real part is still the energy
shift). This is proven elsewhere \cite{[hsfII],[17]}.\smallskip

 The full Hamiltonian to be diagonalized is represented by a $2^N \times
2^N$ matrix. Fortunately, it is block-diagonal in structure,
breaking up into a series of submatrices, in each of which the
coupled subsets of states all have the same number $n$ of excited
qubits.  The submatrices are of dimension $N\choose n$, and the
submatrix for $n$ excited qubits is the same as that for $N-n$
excited, a general property of exchange interactions; this halves
the amount of work we must do (and we consequently tabulate
results only  for
$n=1,\ldots,\lfloor \frac{N}{2}\rfloor$). The
$n=0$ and
$n=N$ eigenstates are just $|G\>$ and $|E\>$ respectively.

\subsection{$n=1:$ Single excitation eigenstates} \label{calcsA}

The single excitation (or $n=1$) eigenstates of a system of  $N$
qubits arranged at the vertices of a regular polygon were
calculated years ago \cite{[hsfIII]}, guided by the symmetry of
the system under rotation about an axis perpendicular to the
polygon plane; the $n=1$ eigenvalues and eigenstates for $N=1-6$
are there tabulated. Here we rewrite these calculations in a
notation which allows us to extend them to states containing
higher numbers of excited qubits, using the case of $N=5$ as an
example.\smallskip

In the subspace spanned by the basis vectors $\{|1\rangle,
|2\rangle,\cdots,|5\rangle\}$, the matrix to be diagonalized has
the form
$$
H_{int}^{(1)}=
\left(
\begin{array}{ccccc}
0 & a & b & b & a\\
a & 0 & a & b & b\\
b & a & 0 & a & b\\
b & b & a & 0 & a\\
a & b & b & a & 0
\end{array}
\right).
$$
 We introduce the matrix $P$, a generator of the 5-dimensional
 representation of $C_5$ (the cyclic group of order 5):
\begin{equation}\label{P}
P=\left(
\begin{array}{ccccc}
0&1&0&0&0\\
0&0&1&0&0\\
0&0&0&1&0\\
0&0&0&0&1\\
1&0&0&0&0
\end{array}
\right).
\end{equation}
The eigenvalue equation of $P$ is given by
$$Pu_{(v)}=\lambda^v u_{(v)},$$ where $\lambda\equiv e^{{2\pi i}\over
5}$, $u_{(v)}=(\lambda^v, \lambda^{2v}, \lambda^{3v},
\lambda^{4v}, \lambda^{5v})$, and $v=1,\cdots,5$. We define the
polynomial $M(x)=a(x+x^4)+b(x^2+x^3)$,
 in terms of which $H_{int}^{(1)}=M(P)$.
Since $H_{int}^{(1)}$ is a sum of powers of $P$, the eigenvectors
$u_{(v)}$ of $P$ will be eigenvectors of $H_{int}^{(1)}$ as well,
and we write the eigenvalue equation
$$H_{int}^{(1)} u_{(v)}=m_{(v)} u_{(v)},$$ where
the eigenvalues $m_{(v)}=M(\lambda^v)\equiv G_{(v)}^{(1)}
+iF_{(v)}^{(1)}$. There is 1 non-degenerate eigenvalue
corresponding to $v=5$, $m_{(5)}=2a+2b$, and (5-1)/2 degenerate
pairs of eigenvalues, corresponding to roots which are complex
conjugates of each other: $\lambda^v=(\lambda^{5-v})^*$. The
eigenvector corresponding to $m_{(5)}$ is simply $u_{(5)}=(1, 1,
1, 1, 1)$. For the eigenvectors corresponding to the degenerate
pairs of eigenvalues, we choose the real linear combinations of
$u_{(v)}$ and $u_{(5-v)}$,
\begin{eqnarray}
RU_{(v)}&=&{1\over 2}(u_{(v)}+u_{(5-v)})\\
 IU_{(v)}&=&{1\over
{2i}}(u_{(v)}-u_{(5-v)}).
\end{eqnarray}
Together with $u_{(5)}$, these form an orthogonal basis set for
the $n=1$ subspace.  They are listed in Table II.

\subsection{$n=2$: Double excitation eigenstates}\label{calcsB}

\subsubsection{Odd values of N}

 We continue with the example of $N=5$ to demonstrate how to calculate the $n=2$
 (biexciton) eigenstates for general odd values of $N$.
 The subspace corresponding to
$N=5$, $n=2$ has 10 basis states, which we take in the order
$\{|12\rangle,|23\rangle,\cdots,|51\rangle;
|13\rangle,|24\rangle,\cdots,|52\rangle\}$.\smallskip

If we define the four polynomials $M_{11}(x)=b(x+x^4)$,
$M_{12}(x)=a(x^4+x^5)+b(x+x^3)$, $M_{21}(x)=a(x+x^5)+b(x^2+x^4)$
and $M_{22}(x)=a(x^2+x^3)$, then the interaction can be
represented by the $10 \times 10$ matrix,
$$H_{int}^{(2)}= M(P)\equiv
\left(\begin{array}{cc}
M_{11}(P)&M_{12}(P)\\
M_{21}(P)&M_{22}(P)
\end{array}
\right).$$ Thus, $H_{int}^{(2)}$ is partitioned  into a $2\times
2$ array of square submatrices, each of dimension $5 \times 5$.
The ability to write the matrix in this form is directly due to
the ordering of the basis vectors, which allows the rotational
symmetry of the pentagon to be reflected in each of the
submatrices.    It is easy to show that for any odd value of $N$,
$H_{int}^{(2)}$ can be partitioned in this way into an array of
$(N-1)/2 \times (N-1)/2$ square submatrices, each of dimension $N
\times N$. This results in a dramatic simplification of the
problem, for instance here we need  diagonalize only a
2-dimensional matrix instead of the original 10-dimensional
one.\smallskip

As with the $n=1$ case discussed above, each matrix $M_{ij}(P)$ is
a linear combination of $P$ and its powers, and therefore has the
eigenvalue equation
$$M_{ij}(P)u_{(v)}=M_{ij}(\lambda^v)u_{(v)},$$
where $\lambda^v$ and $u_{(v)}$ are the eigenvalues and
eigenvectors of $P$. In order to obtain the eigenvalues and
eigenvectors of $H_{int}^{(2)}$, we first solve the eigenvalue
equation
$$M(x)V(x)=\mu(x)V(x),$$  where
$V(x)$ is an eigenvector and $\mu(x)$ an eigenvalue of the
two-dimensional matrix $M(x)$.  The solutions are easily found to
be $$\mu^{\pm}(x)={1\over 2}\big[M_{11}(x)+M_{22}(x)\pm
R(x)\big],$$ where
$$R(x)=\sqrt{\big(M_{11}(x)-M_{22}(x)\big)^2 +
4M_{12}(x)M_{21}(x)},$$ and
$$V^{\pm}(x)=\left(\begin{array}{c}M_{11}(x)-M_{22}(x)\pm
R(x)\\ 2M_{21}(x)\end{array}\right).$$ The eigenvalues and
eigenvectors of $H_{int}^{(2)}$ can then be shown by direct
substitution to be $\{\mu^{\pm}(\lambda^v)\}$ and
$$
U^{\pm}_{(v)}=V^{\pm}(\lambda^v)\otimes u_{(v)}=\left(\begin{array}{c}V_1^{\pm}(\lambda^v)u_{(v)}\\
V_2^{\pm}(\lambda^v)u_{(v)}\end{array}\right),
$$
where $v=1,\cdots,5$. As with $n=1$, for degenerate eigenvalues we
form the real linear combinations of the eigenvectors; the
complete orthogonal  basis set is listed in Table II.\smallskip

 In general, the eigenvalue
equation for the $n=2$ excitations of any odd-$N$ array of qubits
is solved in the same way:\smallskip

\noindent (i) The interaction matrix $H_{int}^{(2)}$ is
partitioned into an array of square submatrices, each of dimension
$N \times N$.\smallskip

\noindent(ii) The eigenvalue equation of matrix $M_{ij}(P)$ is
solved, where $P$ is the $N\times N$ matrix analogous to
Eq.(\ref{P}).\smallskip

\noindent (iii) The eigenvalue equation is solved for the
corresponding $(N-1)/2 \times (N-1)/2$ matrix $M(x)$, yielding
eigenvalues $\{\mu_{(i)}(x)\}$ and eigenvectors
$\{V_{(i)}(x)\}$.\smallskip

\noindent (iv) The eigenvalues and eigenvectors of $H_{int}^{(2)}$
are then given by $\{\mu_{(i)}(\lambda^v)\equiv
G_{(vi)}^{(2)}+iF_{(vi)}^{(2)}\}$ and
$\{U_{(vi)}=V_{(i)}(\lambda^v)\otimes u_{(v)}\}$, where
$v=1,\cdots,N$, $i=1,\cdots,\lfloor(N-1)/2\rfloor$,
$\lambda=e^{\frac{2\pi i}{N}}$, and the vectors $\{u_{(v)}\}$ are
the eigenvectors of the matrix $P$ corresponding to the $N$-sided
polygon.

\subsubsection{Even values of $N$}

 The calculations for the $n=2$ energies and eigenstates for even values of $N$ cannot be
described (or performed) so succinctly. This is due to the fact
that $(N-1)/2$ is an {\it odd} half-integer.  As a result, the
matrix for $H_{int}^{(2)}$ consists of 2 parts: an inner ``core''
of $\lfloor(N-1)/2\rfloor
\times \lfloor(N-1)/2\rfloor$ square submatrices, each of dimension $N
\times N$, plus an outer section of $N/2$ extra columns to the
right and rows at the bottom of the core. For example, the $n=2$
interaction matrix for $N=4$ is given by
$$H_{int}^{(2)}=\left(\begin{array}{cccc|cc}
0&b&0&b&a&a\\
b&0&b&0&a&a\\
0&b&0&b&a&a\\
b&0&b&0&a&a\\
\hline
a&a&a&a&0&0\\
a&a&a&a&0&0\\
\end{array}\right),$$ with an inner core matrix $M(P)=b(P+P^3)$,
 where $P$ is now  the 4-dimensional analogue of
 Eq.(\ref{P}).\smallskip

The calculations are performed in the following manner; we
illustrate the general procedure with the example of
$N=4$:\smallskip

\noindent (i) The energy eigenvalues and vectors of
the ``core'' matrix are obtained, in exactly the same way as
described in the previous section for odd values of $N$.

For the case of $N=4$, the eigenvectors of $M(P)$ are the same as
those of $N=4, n=1$, which in turn are the same as those of $P$.
They appear in Table \ref{table1}.\smallskip

\noindent (ii) These eigenvectors are then divided into 2 groups,
according to their symmetry or antisymmetry.  The vectors $(1, 1,
\cdots, 1, 1)$ and $(1, -1, 1, -1, \cdots, 1, -1)$ are always
eigenvectors, the former symmetric, the latter antisymmetric; the
remainder are classified according to their symmetry under
rotations of $\pi$ about the symmetry axis.

In the case of $N=4$, three of these eigenvectors (those
corresponding to $v=1, 2, \text{ and } 3$ as listed in Table
\ref{table1})  are antisymmetric, while that corresponding to
$v=4$ is symmetric.\smallskip

\noindent(iii) The antisymmetric eigenvectors are appended with $N/2$
$0$'s; the resulting vectors are eigenvectors of $H_{int}^{(2)}$,
and the corresponding energies are found by direct substitution.

In the case of $N=4$, by appending two $0$'s to the ends of the
antisymmetric vectors, we obtain the following three eigenvectors
of
$H_{int}^{(2)}$:
$$U_{(1)}=\left(\begin{array}{c}
1\\ 0\\ -1\\ 0\\ 0\\0\end{array}\right),\; U_{(2)}=\left(\begin{array}{c} -1\\ 1\\ -1\\ 1\\ 0\\
0\end{array}\right), \; U_{(3)}=\left(\begin{array}{c}0\\ 1\\ 0\\ -1\\ 0\\
0\end{array}\right).$$ The corresponding eigenvalues are found by
substitution.  By symmetry, we see that a fourth (antisymmetric)
eigenvector of $H_{int}^{(2)}$ is $U=(0, 0, 0, 0, 1,
-1)$.\smallskip

\noindent(iv) The symmetric eigenvectors are extended into the
rest of the $n=2$ subspace in a symmetric fashion.\smallskip

For the example of $N=4$, the remaining 2 eigenvectors are found
from the symmetric eigenvector $u_{(4)}=(1, 1, 1, 1)$ of
$H_{int}^{(1)}$.  We substitute into the eigenvalue equation for
$H_{int}^{(2)}$ the trial vector $U=(1, 1, 1, 1, x, x)$, obtaining
2 (independent) equations for $x$ and the eigenvalues $\mu$:
$2b+2ax=\mu$ and $4a=\mu x.$ These have the solutions
$\mu^{\pm}=b\pm R,$ $x^{\pm}={4a\over{b\pm R}}$, where
$R=\sqrt{b^2 + 8a^2}$. This completes our set of 6 eigenvectors of
$H_{int}^{(2)}$.  They are listed together with their
corresponding eigenvalues in Table \ref{table1} .\smallskip

We point out that the $N=4$, $n=2$ eigenstates are the first which
depend on the actual strength of the interaction, and not merely
on its symmetry.  In Fig. 2, we illustrate the distance dependence
of their decay rates.  In the long-wavelength limit , three of the
eigenstates have their decay rates unchanged from the
noninteracting value of $2\gamma$, and one state is superradiant,
with an asymptotic value of $5.930\gamma$ (see Table V). The
remaining two states are subradiant:  One shows weak optical
activity, with a limiting decay rate of $0.070\gamma$, and one is
{\it completely} subradiant, with decay rate $\rightarrow0$; thus,
this state is of possible interest for the encoding of quantum
information.  (We have cut off the figure at $\lambda/r=10$ in
order to retain the visibility of some of the oscillations at low
values of the argument, corresponding to shorter wavelengths.)

\subsection{Triple excitation eigenstates}\label{calcsC}

 To complete the sets of
eigenstates for the $N=6$ and $N=7$ polygons, we require those
corresponding to the $n=3$ excitations.  These are obtained with
methods very similar to those used for the $n=2$ states.  For
$N=7$, the $H_{int}^{(3)}$ matrix is first partitioned into a $5
\times 5$ array of square submatrices, each of dimension $7 \times
7$.  The solution then requires the (preliminary) diagonalization
of a $5 \times 5$ matrix $M(P)$, but is otherwise a direct
extension of the method used for $n=2$.\smallskip

For $N=6$ we choose the basis vectors in the order: $\{|123\>,
|234\>,\cdots,|612\>;$ $|124\>,|235\>,\cdots,|613\>;$
$|134\>,|245\>,\cdots,|623\>;$ $|135\>,|246\>\}$. Doing so we find
that the matrix $H_{int}^{(3)}$ consists of a core array of $3
\times 3$ submatrices, each of dimension $6 \times 6$, together
with an outer section of 2 columns to the right and 2 rows at the
bottom of the core.  The solution consists of 2 stages:  In the
first stage, the eigenvalues and eigenvectors of the core matrix
are found, and in the second the symmetric and antisymmetric
eigenvectors of the core are extended to become those of the full
$20 \times 20$ matrix, in a manner entirely analogous to that
employed for $N=6$, $n=2$. The complete set of eigenvalues and
eigenvectors for N=6 is listed in Table III. Those for N=7 are
available upon request.

\section{COOPERATIVE RADIATIVE TRANSITIONS}\label{applications}

In this section, we focus on systems of identical two-level atoms
(or molecular monomers) undergoing cooperative radiative
transitions; the interatomic potential which applies in this case
is the retarded multipole-multipole interaction \cite{[12]}.  The
strongest and most common of these are 1-photon transitions due to
the electric dipole moment operator; however, the same analysis
can also be made for magnetic dipole or (with different
$\Omega_{ij}$) higher electric multipole transitions
\cite{[hsfI],[12],[13]} as well, and even for 2-photon transitions
\cite{[2p]}.  We point out that in many systems an interaction
exists between nearest neighbours (e.g. due to atomic overlap)
{\it in addition} to the electromagnetic exchange interaction
which occurs between all pairs. These forces can be included
trivially in our analysis, simply by incorporating them into the
nearest-neighbour interaction $\Omega_{i,i\pm1}=a$.\smallskip

 For the simple case of linear transition
dipoles, of transition strength $\mu$, oriented (parallel to each
other and) perpendicular to the plane of the ring, $\Omega_{ij}$
can be written in the form
$$\Omega_{ij}=i\gamma\biggl[-{1\over 2}h_2^{(2)}(kr_{ij})+h_0^{(2)}(kr_{ij})\biggr],$$ where
$h_n^{(2)}$ is a spherical Hankel function of the second kind
\cite{[14]} and $\gamma$ is half the atomic Einstein $A$
coefficient,
$$\gamma={{2|\mu|^2\omega^3}\over{3\hbar c^3}}.$$  For linear
transition quadrupoles oriented perpendicular to the plane of the
ring, the interaction is
$$
\Omega_{ij}=i\gamma_q\biggl[-{9\over
28}h_4^{(2)}(kr_{ij})+{5\over 28}h_2^{(2)}(kr_{ij})+{1\over
2}h_0^{(2)}(kr_{ij})\biggr],
$$
where $\gamma_q$ is half the
Einstein $A$ coefficient for the (quadrupole) transition,
$$\gamma_q= {{|q|^2\omega^5}\over {15\hbar c^5}}$$ \cite{[hsfI], [13]}. However,
the analysis is also valid for any system in which the transition
moments of all identical units are oriented symmetrically, i.e.
they form the same angle with the ring. For example, there exist
molecular aggregates in biology known as  ``light-harvesting
complexes'', in which large identical building blocks or monomers
are arranged symmetrically in rings with an N-fold symmetry axis,
whose electronic excitations have been found to extend coherently
over the entire ring \cite{[15]}.  The direction of the transition
moment of each individual monomer forms the (same small) angle
$\theta \ne 90^o$ with the tangent to the ring at its position.
In this case,  the dipole-dipole interaction is (slightly) more
complicated in form, and is given by
$$\Omega_{ij}=i\gamma\bigg[{1\over 2}
\bigl\{3(\hat{\mu}_i\cdot\hat{\bf
r}_{ij})(\hat{\mu}_j\cdot\hat{\bf r}_{ij})
-\hat{\mu}_i\cdot\hat{\mu}_j\bigr\} h_2^{(2)}(kr_{ij}) +(\hat
{\mu}_i\cdot\hat{\mu}_j)\,h_0^{(2)}(kr_{ij})\bigg],$$ but all our
calculations remain valid.\smallskip

 The dynamics of the
system are governed by the Lehmberg-Agarwal master equation
\cite{[16]}, which may be solved by projection onto any complete
set of basis vectors.  However, the ``natural'' set for this
projection are the eigenstates of the Hamiltonian $H=H_f + H_d +
H_{int}$.  As demonstrated previously for $N=2$ and $3$
\cite{[hsfII]} and elsewhere for general $N$ \cite{[17]}, these
states have the following properties:\smallskip

\noindent 1.  The real part of the (complex) eigenvalue gives the
shift of energy of the exciton due to local field
effects.\smallskip

 \noindent 2.  The imaginary part of
the eigenvalue gives the total decay rate or inverse lifetime of
the state; this total decay rate is the sum of the individual
decay rates to all states in the energy manifold below, as
calculated using the master equation and in agreement with the
total energy radiated by the system in a transition between the
two states \cite{[17]}.\smallskip

\noindent 3.  The eigenstates form the basis set within which the
population dynamics and spectroscopic properties of the system are
most conveniently studied.\smallskip

 A calculation of the $n=1$ and
$n=2$ eigenstates of regular polygon systems of odd $N$ for
electric dipole interactions in the long-wavelength limit has in
fact been carried out \cite{[muk]}, but in that reference the
authors included only nearest-neighbour coupling for the real part
of the interaction. For the $n=1$ subspace, the resulting
eigenstates are the same as ours (which however include
interactions between {\it all} neighbours and are valid for
arbitrary wavelength), but the energies  are very different: This
difference is illustrated in Table \ref{table4}, where we list the
$n=1$ energies in the long-wavelength limit for polygons having
$N=5$ and $N=6$, for nearest-neighbour interactions only, for
linear dipoles with all neighbours included, and for linear
quadrupoles with all neighbours included.  (All energies are
expressed in units of the static interaction energy $V_N$ between
a pair of nearest neighbours.) For the $n=2$ subspace, the
eigenstates themselves are very different from those obtained when
only nearest-neighbour interactions are included, and a numerical
comparison of the energies is not meaningful. Because the retarded
interactions are intrinsically long-ranged, a correct calculation
of the eigenstates of the physical system {\it must} include
interactions between all atom pairs:  Only in these states will
the  ``local field'' shifts be the same, and only in these states
will the damping be the same for all atoms, so that no dephasing
occurs during the evolution in time \cite{[18]}. As well, in some
systems the energy of interaction between second (or higher)
nearest neighbours can actually {\it exceed} that between adjacent
pairs (depending on the relative phases of the moments in the
given eigenstate, and/or on the relative orientations of the
transition moments and ${\bf r}_{ij}$).\smallskip

 The detailed emissive properties of these systems will be presented elsewhere \cite{[17]};
however, some simple properties are immediately evident in the
eigenvalues and eigenvectors.  For example, in the long wavelength
limit only one $n=1$ state is optically active in absorption and
emission, and it is superradiant, having an eigenvalue whose
imaginary part $\rightarrow N\gamma$; the $N-1$ other
single-excitation states are subradiant, with the imaginary parts
of their eigenvalues $\rightarrow 0$. In general, the $n=2$ states
decay into $n=1$ states (although these in turn may be
subradiant); however, for even-$N$ polygons, there is (at least)
one $n=2$ state which is itself completely subradiant.\smallskip

 As a simple illustration of the emissive pattern, in Fig. 2 we
display a complete energy level diagram for $N=3$
\cite{[footnote]}. All atomic separations are equal, so that the
system is described by a single interaction potential
$a=\gamma(g+if)$. On each state is indicated its total decay rate,
while individual decay rates between states are indicated on the
dashed lines connecting them. For example, the $n=2$ eigenstate
$(1,\ 1,\ 1)$ has energy $\hbar(2\omega+2\gamma g)$ and a total
decay rate of $2\gamma(1+f)$; it decays at a rate of
$\gamma(4+8f)/3$ to the (symmetric) $(1,\ 1,\ 1)$ state, and at a
rate of $\gamma(1-f)/3$ to each of the 2 antisymmetric states in
the $n=1$ manifold below it.  Similarly, each of the other two
$n=2$ eigenstates has energy $\hbar(2\omega-\gamma g)$ and a total
decay rate of $\gamma(2-f)$. The small sample/long wavelength
limit corresponds to $f\rightarrow 1$.  In this limit, all the
$n=2$ eigenstates decay, but the two antisymmetric $n=1$ states
are (completely) subradiant.\smallskip


\subsection{Absorption from an external field}

If a system in its ground state is placed in a weak external field
of wave vector ${\bf k}$ and polarization $\hat{\bf e}_{\lambda}$,
only the $n=1$ states are excited, with a relative probability
proportional to $|\langle u_{(v)}|\sum_i\,S_i^+\vec{\mu}\cdot
\hat{\bf e}_{\lambda}\,e^{i{\bf k}\cdot{\bf R}_i}|G\rangle|^2$. If
the field is sufficiently intense and the losses sufficiently low,
population can remain in the $\{u_{(v)}\}$ states for long enough
to allow excitation of the $n=2$ states; and so on.\smallskip

In recent years, there has been interest in the excitation of the
exciton and biexciton states of the light-harvesting complexes, in
connection with the calculation of their third-order nonlinear
optical susceptibilities \cite{[muk]}. The complexes discovered so
far have diameters of the order of 10 nm, and their absorption
frequencies correspond typically to wavelengths $\geq 400$ nm, so
that the long-wavelength limit applies.  In this limit, the
dependence on ${\bf k}$ in the absorption probability is
negligible, and it is easy to verify that only those states
$\{u_{(v)}\}$ which are totally symmetric in the atomic positions
are optically active, namely the states $u_{(N)}=(1,\cdots,1)$.
This gives rise to lorentzian (exciton) absorption lines, centred
at the shifted frequencies $G_{(N)}^{(1)}\equiv\omega_0 + \Delta
G_{(N)}^{(1)}$, with (natural) widths
$2F_{(N)}^{(1)}=2N\gamma$.\smallskip

  We denote the
energy of state $U_{(vi)}$ by $G_{(vi)}^{(2)}=2\omega_0 + \Delta
G_{(vi)}^{(2)}$, and its decay constant by $F_{(vi)}^{(2)}$. The
excitation of state $U_{(vi)}$ from a ring in the state $u_{(N)}$
then occurs at frequency $G_{(vi)}^{(2)}-G_{(N)}^{(1)}$.  It can
be shown \cite{[17]} that the width of the absorption line is
$2(F_{(vi)}^{(2)} + F_{(N)}^{(1)})$, and that in the
long-wavelength limit the relative intensities of the biexciton
absorption lines are simply given by $F_{(vi)}^{(2)}$.
\smallskip

In Table V we list the $n=1$ shifts $\Delta G_{(N)}^{(1)}$, the
(unnormalized) $n=2$ eigenvectors $\{U_{(vi)}\}$, and the $n=2$
shifts $\Delta G_{(vi)}^{(2)}$ and decay constants
$F_{(vi)}^{(2)}$ for $N=2,\cdots,9$, in the long-wavelength limit.
The vectors $\{U_{(vi)}\}$ correspond to basis states arranged in
the order $\{|12\rangle,|23\rangle,\cdots,|N1\rangle;
|13\rangle,|24\rangle, \cdots,|N2\rangle;\ etc\}$. In Table VI we
list the corresponding biexciton excitation frequency shifts,
(natural) half widths, and relative intensities.  All frequencies
are expressed in units of $V_N$, and widths in units of $\gamma$.
\bigskip
\section{CONCLUSIONS}

 We have calculated
the eigenstates corresponding to coherent single and multiple
excitations of an array of $N$ identical TLA's or qubits arranged
on the vertices of a regular polygon. Their coupling occurs via an
exchange interaction which depends only on the separation between
the qubits.  We include the interactions between {\it all} pairs,
and our results are valid for arbitrary distances relative to the
radiation wavelength.  To illustrate the usefulness of these
eigenstates, we plot the distance dependence of the decay rates of
the $n=2$ eigenstates of a system of 4 qubits arranged on the
vertices of a square, and tabulate the biexciton eigenstates and
eigenvalues, and absorption frequencies, line widths, and relative
intensities for polygons of $N=2,\cdots,9$ identical TLA's, in the
long-wavelength limit.  The states will be used elsewhere to study
the emissive properties of these systems \cite{[17]} and to
calculate the amount and distribution of entanglement in these
``natural'' entangled states at both zero and finite
temperature.\bigskip

\acknowledgments We wish to thank Tamlyn Rudolph who performed
some of the preliminary calculations. This research was supported
in part by the Natural Sciences and Engineering Research Council
of Canada, the Austrian Science Foundation FWF, the TMR programs
of the European Union, Project No. ERBFMRXCT960087, and by the NSA
\& ARO under contract No. DAAG55-98-C-0040.


\pagebreak
\begin{figure}
\center{\includegraphics[height=40mm]{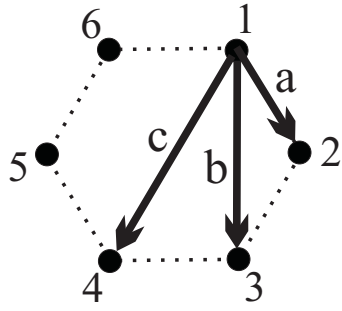}} \caption{A
regular hexagon of interacting qubits}
\end{figure}

\begin{figure}
\center{\includegraphics[height=80mm]{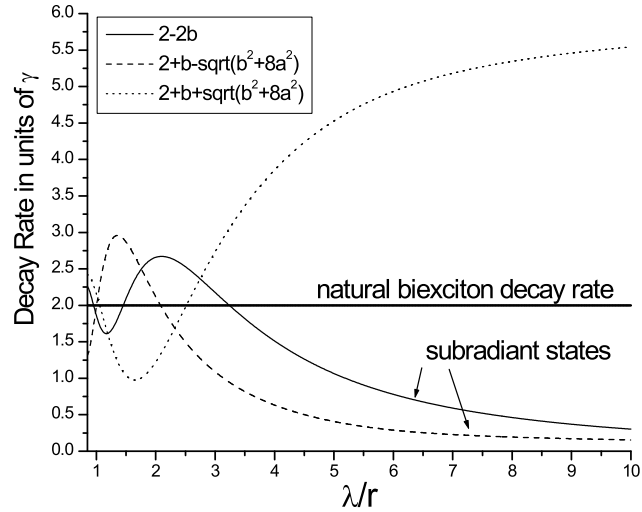}}
\caption{Distance dependence of the decay rates of the $N=4, n=2$
eigenstates}
\end{figure}

\begin{figure}
\center{\includegraphics[height=90mm]{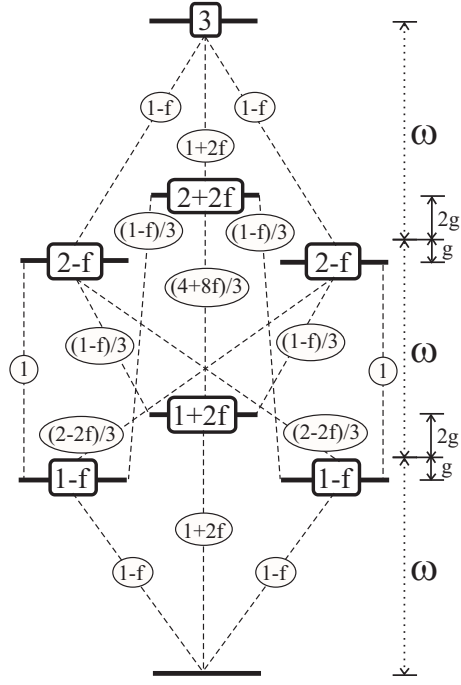}}
\caption{Energy levels and decay rates for an equilateral triangle of qubits (in units of the single qubit decay
rate);
$g$ ($f$) is the real (imaginary) part of the inter-qubit potential $a$.}
\end{figure}

\begin{table}
\begin{tabular}{|c|c|c|c|c|}\toprule
$N$ & $n$ & $v$ & Eigenvalues & Eigenvectors \\\colrule
   2 & 1 & 1 &$ -a$ & $(-1,1)$ \\
   \  & \  & 2 & $a $&
$(1,1)$ \\\colrule
3 & 1 & 1,2 & $-a$ & $(-1,-1,2); (1,-1,0)$ \\
   \  & \  & 3 & $2a$ & $(1,1,1)$ \\\colrule
4 & 1 & 1,3 & $-b$ & $(0,-1,0,1); (1,0,-1,0)$ \\
   \  & \  & 2 &$-2a+b$ &
$(-1,1,-1,1)$ \\
   \  & \  & 4 & $2a+b$ & $(1,1,1,1)$
\\\cline{2-5}
    \  & 2 & 1,3 & $0$ &
$(0,-1,0,1,0,0); (1,0,-1,0,0,0); (0,0,0,0,1,-1) $
\\
  \  & \  & 2 & $-2b$ & $(-1,1,-1,1,0,0)$ \\
    \
& \  & 4 & $b\pm R$ & (1,1,1,1,$x_{\pm},x_{\pm}$)
\\\cline{4-5}
\ & \  & \  & \multicolumn{2}{c|}{$R=\sqrt{b^2+8a^2}$ \;\; $x_{\pm}=(-b\pm R)/2a$} \\
\botrule
\end{tabular}
\caption{Eigenvalues and eigenvectors of a single pair (N=2), equilateral
triangle (N=3), and square array (N=4) of qubits.}\label{table1}
\end{table}

\begin{table}
\begin{tabular}{|c|c|c|c|} \toprule %
  $n$ & $v$ & Eigenvalues & Eigenvectors \\\colrule
     1 & 1,4
& $2(c_2 a-c_1 b)$ & $(c_2,c_4,c_6,c_8,c_{10});
(s_2,s_4,s_6,s_8,s_{10})$ \\
   \  & 2,3 & $2(-c_1 a+c_2 b)$ & $(c_4,c_8,c_{12},c_{16},c_{20}); (s_4,s_8,s_{12},s_{16},s_{20})$\\
   \  & 5 & $2a+2b$ & (1,1,1,1,1) \\\colrule
   2 & 1,4 & $E_{\pm}(-c_1 a,c_2 b)$ & $(c_2,-c_1,-c_1,c_2,1;v_{\pm}c_2,v_{\pm},v_{\pm}c_2,-v_{\pm}c_1,-v_{\pm}c_1)$
\\
   \ & \  & \  & $(-1,-2c_2,2c_2,1,0;v_{\pm},0,-v_{\pm},-2v_{\pm}c_2,2v_{\pm}c_2)$
\\
   \  & 2,3 & $E_{\pm}(c_2 a,-c_1 b)$ &
$(-c_1,c_2,c_2,-c_1,1;-w_{\pm}c_1,w_{\pm},-w_{\pm}c_1,w_{\pm}c_2,w_{\pm}c_2)$
\\
   \  & \  & \  &
$(-2c_2,1,-1,2c_2,0;2w_{\pm}c_2,0,-2w_{\pm}c_2,w_{\pm},-w_{\pm})$
\\
   \  & 5 & $E_{+}(a,b)$ &
$(1,1,1,1,1;u,u,u,u,u)$ \\
   \  & \  & $E_{-}(a,b)$ &
$(u,u,u,u,u;-1,-1,-1,-1,-1)$\\\cline{2-4}
&  \multicolumn{3}{c|}{$\begin{array}{rclrcl}
c_j&=&\cos(j\pi/5)& u&=&G_{+}(a,b)\\
s_j&=&\sin(j\pi/5)& v_{\pm}&=&G_{\pm}(-c_1 a,c_2 b)\\
F(\alpha,\beta)&=&\sqrt{5(\alpha+\beta)^2 -4\alpha \beta}&w_{\pm}&=&G_{\pm}(c_2 a, -c_1 b)\\
E_{\pm}(\alpha,\beta)&=&\alpha+\beta\pm
F(\alpha,\beta)&G_{\pm}(\alpha,\beta)&=&[\alpha-\beta\pm
F(\alpha,\beta)]/2(\alpha+\beta)
\end{array}$}

\\
\botrule
\end{tabular}
\caption{Eigenvalues and eigenvectors of 5 qubits arranged on the vertices of a regular pentagon.}\label{table2}
\end{table}

\begin{table}
\begin{tabular}{|c|c|c|c|}\toprule
  $n$ & $v$& Eigenvalues & Eigenvectors \\\colrule
  1 & 1,5 & $a-b-c$ & $(1,-1,-2,,-1,,1,2);(1,1,0,-1,-1,0)$ \\   \  & 2,4 &
$-a-b+c$ & $(-1,-1,2,-1,-1,2); (1,-1,0,1,-1,0)$ \\   \  & 3 &
$-2a+2b-c$ & $(-1,1,-1,1,-1,1)$ \\   \  & 6 & $2a+2b+c$ &
$(1,1,1,1,1,1)$ \\\colrule
   2 & 1,5 & $\pm
\sqrt{b^2 +3a^2}$ & $(1,-1,-2,-1,1,2;0,-u_{\pm},-u_{\pm},0,u_{\pm},u_{\pm};0,0,0)$  \\
\  & \  & \  & $(3,3,0,-3,-3,0;2u_{\pm},u_{\pm},-u_{\pm},-2u_{\pm},-u_{\pm},u_{\pm};0,0,0)$ \\
\  & 2,4 & $\nu$ &
$(-1,-1,2,-1,-1,2;-2x,x,x,-2x,x,x;-y,2y,-y)$ \\   \  &  &  &
$(1,-1,0,1,-1,0;0,-x,x,0,-x,x;-y,0,y)$ \\   \  & 3 & $-2b$ &
$(-1,1,-1,1,-1,1;0,0,0,0,0,0;0,0,0)$ \\   \  & \  &$2b$  &
$(0,0,0,0,0,0;-1,1,-1,1,-1,1;0,0,0)$ \\   \  & 6 & $\mu$ &
$(1,1,1,1,1,1;r,r,r,r,r,r;s,s,s)$ \\\cline{2-4}
& &
\multicolumn{2}{c|}{$\small\begin{array}{c}\begin{array}{rclrcl}
\\[-8pt]
\nu^3 + 2b\nu^2 +(4ac-b^2-3a^2-4c^2)\nu+2b(b^2-a^2+4ac)&=&0  &\;\; x&=&(\nu^2+b\nu-2b^2)/[\nu(a-2c)-2ab]\\
\mu^3-4b\mu^2-4(2ac+b^2+3a^2+c^2)\mu+16b(b^2-a^2-2ac)&=&0
&\;\;y&=&2(a\nu+2bc)/[\nu(a-2c)-2ab]\\\end{array} \\
r=(\mu^2-2b\mu-8b^2)/2[\mu(a+c)+4ab]\;\;\;\;\;\;
s=2(a\mu+2bc)/[\mu(a+c)+4ab]\;\;\;\;\;\; u_{\pm}=-b \pm
\sqrt{b^2+3a^2}
\end{array}$}\\\colrule
3 & 1,5 & $a-b-c+(a+b)m_{\pm}$ &
$(m_{\pm},-m_{\pm},-2m_{\pm},-m_{\pm},m_{\pm},2m_{\pm};1,-1,-2,-1,1,2;-1,-2,-1,1,2,1;0,0)$
\\   \  & \  & \  &
$(m_{\pm},m_{\pm},0,-m_{\pm},-m_{\pm},0;1,1,0,-1,-1,0;1,0,-1,-1,0,1;0,0)$
\\   \  & $1,5\atop 2,4$ & $b\mp a\mp c$ &
$(0,0,0,0,0,0;-2,\mp1,1,\pm2,1,\mp1;\pm1,-1,\mp2,-1,\pm1,2;0,0)$
\\   \  & \  & \  &
$(0,0,0,0,0,0;0,-1,\mp1,0,\pm1,1;1,\pm1,0,\mp1,-1,0;0,0)$ \\   \
& 2,4 & $c-a-b+(a-b)n_{\pm}$ &
$(n_{\pm},n_{\pm},-2n_{\pm},n_{\pm},n_{\pm},-2n_{\pm};1,1,-2,1,1,-2;1,-2,1,1,-2,1;0,0)$
\\   \  & \  & \  &
$(n_{\pm},-n_{\pm},0,n_{\pm},-n_{\pm},0;1,-1,0,1,-1,0;-1,0,1,-1,0,1;0,0)$
\\   \  & 3 & $\sigma_{-}$ &
$(-p_{-},p_{-},-p_{-},p_{-},-p_{-},p_{-};1,-1,1,-1,1,-1;-1,1,-1,1,-1,1;-q_{-},q_{-})$
\\   \  & $3\atop 6$ & $\mp c\pm 2a-2b$ & $(0,0,0,0,0,0;\mp
1,1,\mp 1,1,\mp 1,1;-1,\pm 1,-1,\pm 1,-1,\pm 1;0,0)$ \\
\  & 6 & $\sigma_{+}$ &
$(p_{+},p_{+},p_{+},p_{+},p_{+},p_{+};1,1,1,1,1,1;1,1,1,1,1,1;q_{+},q_{+})$\\
\cline{2-4}
& & \multicolumn{2}{c|}{
$\begin{array}{rcl}
\\[-8pt]
\sigma_{\pm}^3&-&(2b\pm 3c\pm
2a)\sigma_{\pm}^2-[4(a^2 +b^2)+(2a\pm 2b-c)^2]\sigma_{\pm}\pm 3c[c^2 +2(\pm bc \mp 4ab +ac)]=0\\
p_{\pm} &=& 2[\left(2b \pm a\right)\sigma_{\pm} + 3ca]/[\sigma^2_{\pm}\mp 2c\sigma_{\pm}- 3c^2]\\
q_{\pm} &=& 6[a\sigma_{\pm} \mp ac +2bc]/[\sigma^2_{\pm}\mp 2c\sigma_{\pm}-3c^2]\\
m_{\pm}&=&[{{2c+b-a\pm\sqrt{(2c+b-a)^2+8(a+b)^2}}]/[{2(a+b)}}]\\
n_{\pm}&=&[{{a+b-2c\pm\sqrt{(2c-b-a)^2+8(a-b)^2}}]/[{2(a-b)}}] \end{array}$}\\
\botrule
\end{tabular}
\caption{Eigenvalues and eigenvectors of 6 qubits arranged on the vertices of a
regular hexagon.}\label{table3}
\end{table}
\begin{table}
\begin{tabular}{|c|c|c|c|c|c|}\toprule
$N$  & $v$ & Energy & nearest neighbours & linear dipoles, &
linear quadrupoles, \\   \ \ \  & \  & \ \  & only & all
neighbours & all
neighbours \\\hline 5& 1, 4 & $.618a-1.618b$ &.618 & .239 & .474\\
\ \ & 2, 3& $-1.618a+.618b$ & $-1.618$& $-1.473$ & $-1.563$ \\   \
\  & 5 & $2a+2b$   &       2 &   2.468 & 2.178
\\\hline    6 & 1, 5 & $a-b-c$ &  1&   .683& .905\\   \ \ & 2, 4
& $-a-b+c$ & $-1$ & $-1.067$ & $-1.033$\\   \ \ & 3 & $-2a+2b-c$&
$-2$&  $-1.741$& $-1.902$ \\ \ \ \  & 6 & $ 2a+2b+c$ & 2 & 2.509 &
2.159\\\hline \botrule
\end{tabular}
\caption{$n=1$ energies in the small sample limit (units of the
static interaction energy between nearest
neighbours).}\label{table4}

\end{table}

\begin{table}
\begin{tabular}{|c|c|c|c|c|}\toprule
  $N$ & $\Delta G_{(N)}^{(1)}$ & $n=2$ eigenvectors & $\Delta G_{(vi)}^{(2)}$ & $F_{(vi)}^{(2)}$ \\
  \  & (units of $V_N$) & \  & (units of $V_N$) & (units of $\gamma$) \\\hline
  2 & 1 & $(1)$ & 0 & 2 \\\hline
  3 & 2 & $(1,1,1)$ & 2 & 4 \\\hline
  4 & 2.354 & $(1,1,1,1,x,x)$ & \  & \  \\
  \  & \  & $x=1.248-.283i/V_N$ & 3.204 & 5.930 \\
  \  & \  & $x=-1.602-.363i/V_N$ & -2.496 & .070 \\\hline
  5 & 2.472 & $(1,1,1,1,1;x,x,x,x,x)$ & \  & \  \\
  \  & \  & $x=1.356+.649i/V_N$ & 3.823 & 7.821 \\
  \  & \  & $x=-.737-.352i/V_N$ & -1.351 & .179 \\\hline
  6 & 2.511 & $(1,1,1,1,1,1;x,x,x,x,x,x;y,y,y)$ & \  & \  \\
  \  & \  & $x=1.417-1.261i/V_N; \ y=1.548-1.732i/V_N$ & 4.162 & 9.687 \\
  \  & \  & $x=-.114-.373i/V_N; \ y=-1.087-.550i/V_N$ & -.290 & .293 \\
  \  & \  & $x=-1.940-.312i/V_N;\  y=2.252+.200i/V_N$ & -3.100 & .020 \\\hline
  7 & 2.518 & $(1,1,1,1,1,1,1;x,x,x,x,x,x,x;y,y,y,y,y,y,y)$ & \  & \  \\
  \  & \  & $x=1.45-1.602i/V_N;\  y=1.66-2.738i/V_N$ & 4.358 & 11.534 \\
  \  & \  & $x=.314-.433i/V_N;\  y=-.877-.765i/V_N$ & .570 & .410 \\
  \  & \  & $x=-1.416-.379i/V_N;\  y=.634+.011i/V_N$ & -2.410 & .056 \\\hline
  8 & 2.515 & $(x,x,x,x,x,x,x,x;1,1,1,1,1,1,1,1;y,y,y,y,y,y,y,y;z,z,z,z)$ & \  & \  \\
  \  & \  & $x=0+59.68i/V_N;\  y=1.177-.949i/V_N;\  z=1.237-1.444i/V_N$ & 4.478 & 13.37 \\
  \  & \  & $x=1.635+69.72i/V_N;\  y=0-.0714i/V_N;\  z=-1.809-132.87i/V_N$ & 1.248 & .528 \\
  \  & \  & $x=-1.0914-285.8i/V_N;\  y=.4523+.0204i/V_N;\  z=-1.284+.453i/V_N$ & -1.660 & .097 \\
  \  & \  & $x=-.490-63.44i/V_N;\  y=-1.33-43.48i/V_N;\  z=1.451-34.91i/V_N$ & -3.322 & .008 \\\hline
  9 & 2.508 & $(1,1,1,1,1,1,1,1,1;x,x,x,x,x,x,x,x,x;y,y,y,y,y,y,y,y,y;z,z,z,z,z,z,z,z,z)$ & \  & \  \\
  \  & \  & $x=1.493-2.834i/V_N;\  y=1.793-5.665i/V_N;\  z=1.939-7.341i/V_N$ & 4.559 & 15.19 \\
  \  & \  & $x=.822-.583i/V_N;\  y=-.144-1.184i/V_N;\  z=-1.015-1.518i/V_N$ & 1.780 & .644 \\
  \  & \  & $x=-.496-.429i/V_N;\  y=-.916-.396i/V_N;\  z=.713-.0043i/V_N$ & -.958 & .140 \\
  \  & \  & $x=-1.713-.386i/V_N;\  y=1.538+.177i/V_N;\  z=-.619-.222i/V_N$ & -2.874 & .025 \\\hline
\botrule

\end{tabular}
\caption{Frequency shifts and widths of the exciton and biexciton
energy levels (long-wavelength limit).}
\end{table}

\begin{table}
\begin{tabular}{|c|c|c|c|}\toprule
  $N$ & Frequency shifts & Half-widths & Relative Intensities \\
  \  & (units of $V_N$) & (units of $\gamma$) & \  \\\hline
  2 & $-1$ & 4 & 1 \\\hline
  3 & 0 & 7 & 1 \\\hline
  4 & 0.85 & 9.93 & .988 \\
  \  & $-4.85$ & 4.07 & .012 \\\hline
  5 & 1.351 & 12.821 & .978 \\
  \  & $-3.823$ & 5.179 & .022 \\\hline
  6 & 1.651 & 15.687 & .969 \\
  \  & $-2.801$ & 6.293 & .029 \\
  \  & $-5.611$ & 6.020 & .002 \\\hline
  7 & 1.840 & 18.534 & .961 \\
  \  & $-1.948$ & 7.410 & .034 \\
  \  & $-4.929$ & 7.056 & .005 \\\hline
  8 & 1.964 & 21.370 & .9550 \\
  \  & $-1.266$ & 8.528 & .0377 \\
  \  & $-4.174$ & 8.097 & .0069 \\
  \ & $-5.836$ & 8.008 & .0006 \\\hline
  9 & 2.052 & 24.190 & .9494 \\
  \  & $-.727$ & 9.644 & .0403 \\
  \  & $-3.465$ & 9.140 & .0088 \\
  \  & $-5.381$ & 9.025 & .0016\\\hline
\botrule
\end{tabular}
\caption{Biexciton excitation frequency shifts, natural line
widths, and relative intensities (long-wavelength limit).}
\label{table2}
\end{table}

\end{document}